%% file: teranet_paper.tex
\documentclass[conference]{IEEEtran}
\input{input.tex}
\usepackage{dblfloatfix} 
\usepackage{epstopdf}
\usepackage{enumerate}
\usepackage{algorithm}
\usepackage{amsmath}
\usepackage{amssymb,amsfonts}
\usepackage[noend]{algpseudocode}
\usepackage{float}
\usepackage{color}
\usepackage{makeidx}
\usepackage{bbm}
\usepackage{lipsum}
\usepackage{balance}
\usepackage{wrapfig}

\usepackage{url}
\IEEEoverridecommandlockouts
\usepackage{graphicx}

\usepackage{cleveref}
\usepackage{steinmetz}
\usepackage{varwidth}
\usepackage{textcomp}
\usepackage[usenames,dvipsnames]{xcolor}

\usepackage{caption}
\usepackage{color}

\algnewcommand{\Initialize}[1]{%
	\State \textbf{Initialization:} \parbox[t]{.8\linewidth}{\raggedright #1}}

\begin{document}
	\title{Deep Learning for THz Drones with Flying Intelligent Surfaces: Beam and Handoff Prediction\thanks{This paper is based upon work funded by AFRL/SBIR program under AFRL contract No. FA8750-20-C-0505.}}
	\author{Nof Abuzainab$^1$, Muhammad Alrabeiah$^2$, Ahmed Alkhateeb$^2$, and Yalin E. Sagduyu$^1$
	\\$^1$ \textit{Intelligent Automation, Inc.}, email: $\{$nabuzainab, ysagduyu$\}$@i-a-i.com\\ 
	$^2$ \textit{Arizona State University}, emails:  $\{$malrabei, alkhateeb$\}$@asu.edu\\
	}
	\maketitle

\begin{abstract}
We consider the problem of proactive handoff and beam selection in Terahertz (THz) drone communication networks assisted with reconfigurable intelligent surfaces (RIS). Drones have emerged as critical assets for next-generation wireless networks to provide seamless connectivity and extend the coverage, and can largely benefit from operating in the THz band to achieve high data  rates (such as considered for 6G). However, THz communications are highly susceptible to channel impairments and blockage effects that become extra challenging when accounting for drone mobility. RISs offer flexibility to extend coverage by adapting to channel dynamics. To integrate RISs into THz drone communications, we propose a novel deep learning solution based on a recurrent neural network, namely the Gated Recurrent Unit (GRU), that proactively predicts the serving base station/RIS and the serving beam for each drone based on the prior observations of drone location/beam trajectories. This solution has the potential to extend the coverage of drones and enhance the reliability of next-generation wireless communications. Predicting future beams based on the drone beam/position trajectory significantly reduces the beam training overhead and its associated latency, and thus emerges as a viable solution to serve time-critical applications. Numerical results based on realistic 3D ray-tracing simulations show that the proposed deep learning solution is promising for future RIS-assisted THz networks by achieving near-optimal proactive hand-off performance and more than 90$\%$ accuracy for beam prediction. 
\end{abstract}

\section{Introduction}
With the unprecedented increase in the number of devices (e.g., IoT devices) and new applications (e.g., virtual/augmented reality) that require wireless connectivity \cite{IoT}, there is a critical need to redesign wireless networks that can support the ever-growing demand to improve the communication rates. 
Due to the abundance of bandwidth at the Terahertz (THz) band (ranging between 0.1 THz and 10 THz) \cite{THz1,THz2}, THz communication is envisioned to meet the rate demands of next-generation wireless communication networks (in particular, 6G \cite{6G}) by pushing the rates from the Gigabits  per second (Gbps) to the Terabits per second (Tbps). One potential application of THz communications is in airborne networks, where unmanned aerial vehicles (UAVs) or so called drones form a network to extend the connectivity and coverage for communication and surveillance needs, e.g., collectively transferring the overhead imagery that they take to a ground base station. 
However, there are several challenges to overcome before realizing the anticipated benefits of THz communications. The THz channel suffers from severe path loss and molecular absorption, which limits significantly the communication distance on the THz band. In addition, due to the short wavelength of the THz signal, communication is hindered by most of the obstacles present along the communication path, including human users. High network mobility such as due to the mobility of drones aggregates the impact of this challenge. Thus, novel solutions are required that are aware of blockages and extend the range of the THz communication, thereby increasing its feasibility and scale. Among these promising solutions are the use of highly directional antennas, adaptive beamforming, and multi-hop relaying \cite{THz2}.

Another promising solution to expand the coverage and range of THz communication is the use of the reconfigurable intelligent surface (RIS) \cite{LIS1, LIS2, LIS3}. As an emerging technology to move from massive antennas to antenna surfaces for software-defined wireless systems, the RIS relies on an array of unit cells/elements to control the scattering and reflection profiles of the electromagnetic signals, allowing considerable mitigation of the propagation loss and multipath attenuation. The RIS is very promising, in particular, for THz communication that is highly sensitive to pathloss and blockage. Also, the RIS can be used for effective beamforming of the THz signals as it is still challenging to realize digital beamforming at the THz band.

\subsection{Prior Work}
Beam selection has been considered in mmWave systems using deep learning techniques \cite{Cousik, Cousik2} and base station selection has been incorporated in \cite{sysmodel}. Deep learning has been also applied to beam selection for the RIS \cite{LIS4}. 
There have been recent works that consider incorporating RISs into THz communication networks. The deployment of the RIS was considered for indoor THz communication scenarios in \cite{LISindoor}. The application of the holographic RIS in the THz band was considered in \cite{LISholographic}. Closed-loop and compressed sensing channel estimation algorithms were proposed to estimate the THz channels. In \cite{LISsatelite}, communication over the THz band was considered for an inter-satellite network to achieve high data rate transmissions, and the gains of using the RIS on neighboring satellites were assessed in terms of error rate. The use of RISs was considered in \cite{LISbeamforming} and \cite{LISopt} to assist in serving mobile users that are non line-of-sight with the serving base station. In \cite{LISbeamforming}, cooperative beamtraining techniques were proposed to estimate the channel with the RIS and low-cost beamforming algorithms were then proposed based on the developed channel estimation. The problem of optimizing the phase shift of the RIS was considered in \cite{LISopt} to serve a set of users that are non line-of-sight (NLOS) with the serving base station, communicating over the THz band. 

However, none of these works considered the mobility of users or RISs for THz communications. Mobility increases the challenges for THz communications. This is due to the fact that THz communications is extremely sensitive to mobility due the high directionality of THz links. Furthermore, due to the ultra-high data rates of the THz links, any disruption in communications will result in queue overflow and severe data loss, which will consequently lead to significant service degradation. Thus, novel proactive solutions that maintain the communication links are essential to realize reliable communications in the THz band.

\subsection{Contribution}
To tackle the aforementioned challenges, we consider a THz drone network in which a mobile drone user is served by a base station and a flying RIS. We address two main problems in this scenario: (i) prediction of the optimal beamforming vector for the base station and the RIS to serve the mobile drone, and (ii) drone proactive hand-off from the base station to the RIS and vice versa. As deep learning has been shown to effectively learn from and adapt to spectrum data \cite{deeplearning}, we consider a deep neural network solution for beam prediction and proactive hand-off. We summarize our contribution in the following points:
\begin{itemize}
\item We propose an RIS-assisted drone THz-communication network, in which a mobile drone maintains a consistent communication link with the base station either through a direct link or an RIS-assisted link.
\item We formulate a deep learning algorithm to realize the consistent communication link between a base station and a drone. The algorithm proactively predicts the best communication link (direct or RIS-assisted) and predicts the best beamforming vector for that link. As we account for drone mobility, we consider a recurrent neural network solution based on Gated Recurrent Units (GRUs) to capture temporal correlations in the spectrum data and learn the sequence dependency.  
\item We have built a dataset to study and evaluate the proposed algorithm. First, We show that the proposed deep learning algorithm achieves high accuracy in determining the optimal communication link as well as the serving beams. Then, we show that a minor form of beam training can further increase the accuracy of the proposed deep learning algorithm reaching up to $11\%$ when top-3 accuracy is considered for beam prediction as opposed to top-1 accuracy.
\end{itemize}

The rest of the paper is organized as follows. Section~\ref{sec:sys_model} presents the system and the channel model. Section \ref{sec:prob_def} 
formulates the problem. Section~\ref{sec:DL} presents the proposed deep learning solution. Section~\ref{sec:exp_setup} describes the dataset for performance evaluation. Section~\ref{sec:exp_results} presents the numerical results. Section~\ref{sec:conclusion} concludes the paper.

\section{System and Channel model}\label{sec:sys_model}

\begin{figure}[t]{
	\centering
	\includegraphics[width=9 cm,height=4.5cm,angle=0]{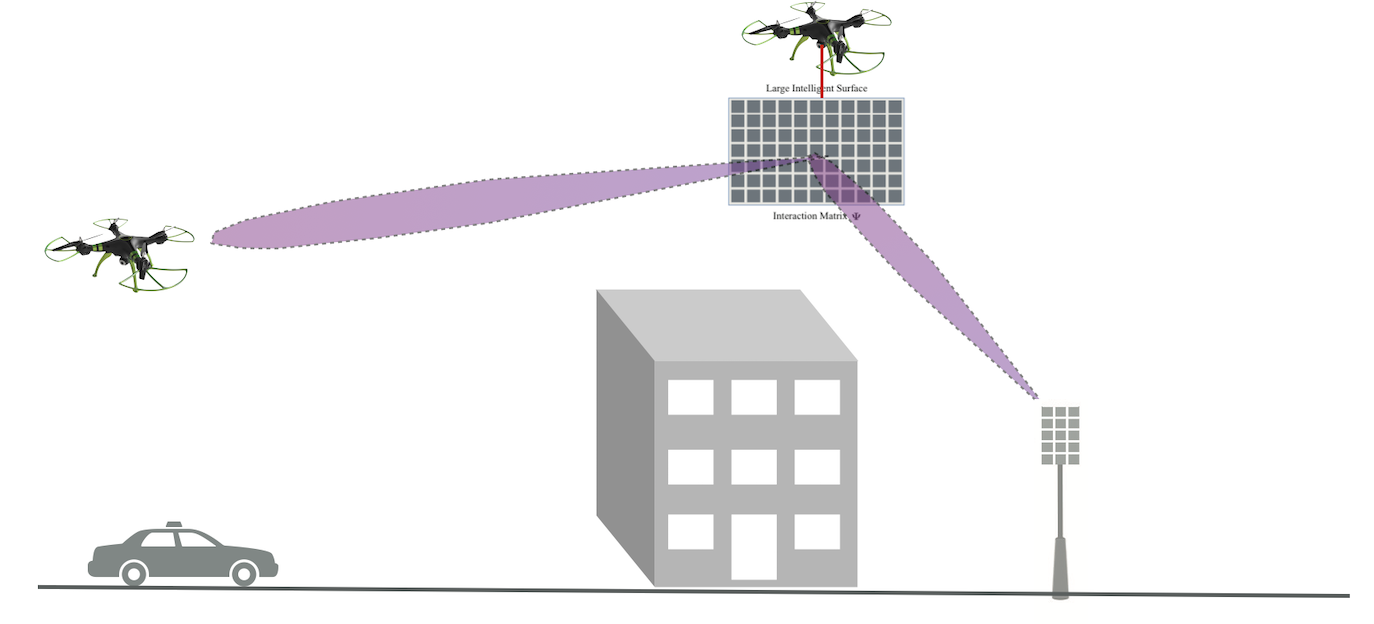}
	\caption{Outdoor drone scenario with the RIS.}\label{fig:sys_model}
	}
\end{figure}

\begin{figure*}[t!]
	\centering
	\includegraphics[width=\linewidth]{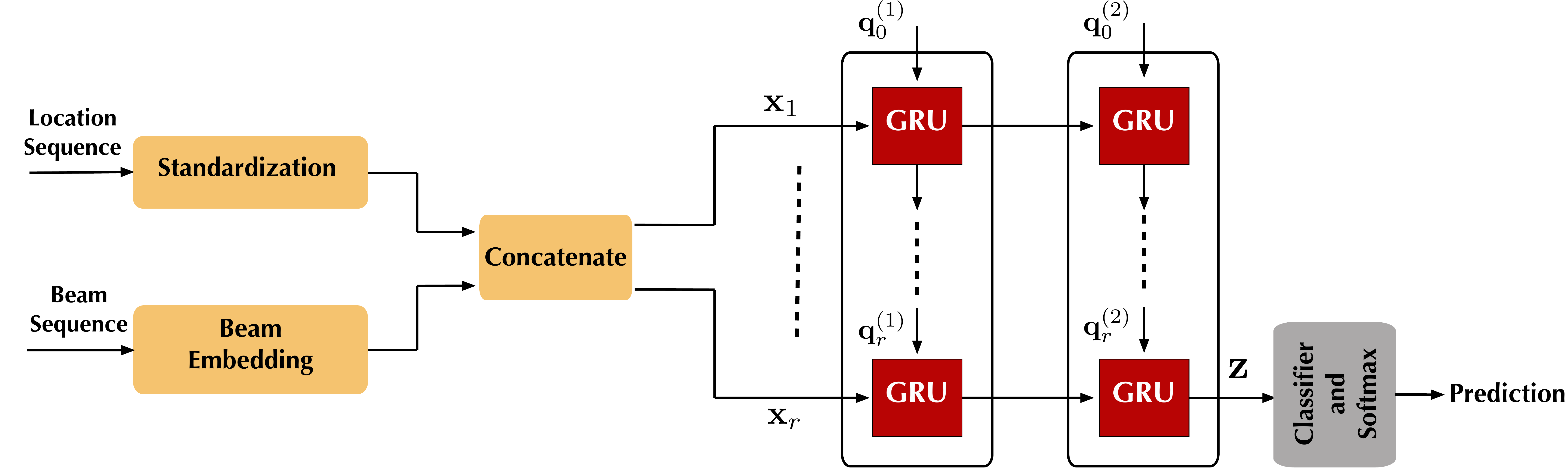}
	\caption{Deep learning network structure to predict the serving base station and RIS.}\label{fig:deep_arch}
\end{figure*}

The THz system adopted in this paper is depicted in Figure \ref{fig:sys_model}. It has a single base station equipped with $M$-element antenna array that serves a mobile drone user with a beamforming vector $\bff \in \mathbb{C}^{M\times1}$. The beamforming vectors are selected from a predefined beam codebook  $\mathcal F$ of size $M_\mathrm{CB}$. If the LOS link between the base station and drone is lost, the base station serves the drone through a flying intelligent surface, equipped with $N$ antennas. Let $\mathbf{h}_{T,k}$ and $\mathbf{h}_{R,k}$ denote the uplink channel matrices of the transmitter and receiver to the intelligent surface, and let $\mathbf{h}^T_{T,k}$ and $\mathbf{h}^{T}_{R,k}$ denote the downlink channel matrices of the transmitter and receiver to the intelligent surface. The received signal strength at the $k$-th subcarrier is expressed as
\begin{eqnarray}
y_k&=&\mathbf{h}^{T}_{R,k}\mathbf{\Psi} \mathbf{h}_{T,k}s+v\\
&=&(\mathbf{h}^{T}_{R,k} \odot \mathbf{h}_{T,k})^T \psi s +v, \label{lisreceivedsignal}
\end{eqnarray}
where $s\in\mathbb C$ is a data symbol satisfying $\mathbb E\left[|s|^2\right]=P$, $P$ is the total transmit power, $v\sim\mathcal N_{\mathbb C}(0,\sigma^2)$, and $\mathbf{\Psi} \in\mathbb C^{I \times I}$ is the RIS interaction matrix and represents the interaction of the RIS with the incident signal from the transmitter. Note that (\ref{lisreceivedsignal}) is a result of the diagonal structure of the interaction matrix $\mathbf{\Psi}$, i.e., $\mathbf{\Psi}=diag(\psi)$ where $\psi$ is the diagonal vector.
This diagonal structure results from the operation of the RIS where every element $i$, $i = 1, 2, ...,I$ reflects the incident signal after multiplying it with an interaction factor $[\psi]_i$. The interaction factor $[\psi]_i$ is given by $[\psi]_i=e^{j\phi_i}$ considering that the RIS elements are implemented using phase shifters only. Further, we will call the interaction vector  
in this case the reflection beamforming vector. The reflection beamforming vector $\psi$ is selected from a reflection beamforming codebook $\mathcal{P}$.

\textbf{Channel Model:}
We adopt a wideband geometric THz channel model \cite{THz2} with $L$ clusters. Each cluster $\ell$, $\ell =\{1,...,L\}$ is assumed to contribute with one ray that has a time delay $\tau_\ell \in \mathbb{R}$,  azimuth/elevation angles of arrival (AoA) represented by $(\theta_\ell$, $\phi_\ell)$, and complex path gain $\alpha_\ell$ (which includes the path loss). Further, let $p_\mathrm{rc}(\tau)$ represent a pulse shaping function for T-spaced signaling evaluated at seconds. 
With this model, the delay-d channel between the user and the  base station follows
\begin{equation}
\boldsymbol{\mathsf{h}}_d=\sum_{\ell=1}^L {\rho} \alpha_l p_\mathrm{rc}(dT_s-\tau_\ell)a_{\text{rv}}(\theta_\ell,\phi_\ell), \label{delay}
\end{equation}
where $a_{\text{rv}}(\theta_\ell,\phi_\ell)$ is the array response vector of the base station at the AoAs $(\theta_\ell$,$\phi_\ell)$. Given the delay-d channel in (\ref{delay}), the frequency domain channel vector at subcarrier $k$, $\mathbf{h}_{k}$, can be written as
\begin{equation}
\mathbf{h}_{k}=\sum_{d=0}^{D-1}\boldsymbol{\mathsf{h}}_d \ e^{-j\frac{2\pi k}{K}d}.
\end{equation}
Considering a block-fading channel model, $\{\mathbf{h}_{n,k}\}_{k=1}^K$  are assumed to stay constant over the channel coherence time, denoted by $T_C$ \cite{coherencetime1}.

\section{Problem Definition and Formulation}\label{sec:prob_def}
The objective is to predict the best communication link and its serving beam. To formulate the two problems, we define the following.

\begin{itemize}
	\item \textbf{Beam Sequence:} Due to the mobility of the user, the serving base station frequently updates its beam 
	$\mathbf{f}$ every time instant corresponding to beam coherence time. This beam coherence time depends on many factors including the speed of the user and the number of antennas at the base station. To account for the beam coherence time, we define $\mathbf{f}^{(t)}$ as the beam used by the base station to serve the mobile drone at beam coherence time $t$, where $t = 1, 2, ...$ and with $t = 1$ representing the first beam coherence time at which the drone is first connected to the base station. Based on this, we define a $t$-step sequence of beams as
	\begin{equation}
	\mathcal{B}_t=\{\mathbf{f}^{(1)}, \mathbf{f}^{(2)},...,\mathbf{f}^{(t)}\}.
	\end{equation}
	\item \textbf{Position Sequence:} Let $\mathbf{x}^{(t)}$ denote the position at time step $t$ (i.e., when beam $\mathbf{f}^{(t)}$ was selected). Then, we define a t-step sequence of positions as
	\begin{equation}
	\mathcal{X}_t=\{\mathbf{x}^{(1)}, \mathbf{x}^{(2)},...,\mathbf{x}^{(t)}\}.
	\end{equation}
	\item \textbf{LIS Index Sequence:} Let $\psi^{(t)}$ denote the reflection beamforming vector selected at time step $t$. We define a t-step sequence of RIS beams as
	\begin{equation}
	\mathcal{L}_t=\{\psi^{(1)}, \psi^{(2)},...,\psi^{(t)}\}.
	\end{equation} 
	\item \textbf{Communication-link Sequence:} Let $b^{(t)}$ denote the indicator of whether the base station has a direct link to the mobile drone at time $t$, or not. We define a t-step sequence of communication links as
	\begin{equation}
	\mathcal{W}_t=\{b^{(1)},b^{(2)},...,b^{(t)}\}.
	\end{equation}
\end{itemize}

Using the above definitions, we formulate our problem as predicting the communication link and the serving beam at time instance $t + 1$ given the sequence of beams ($\mathcal L_t$ and $\mathcal B_t$) and positions ($\mathcal X_t$).
Formally, we design a machine learning algorithm to learn the mapping $ \left\{ \mathcal B_t, \mathcal{L}_t,  \mathcal{W}_t \right\} \rightarrow \left\{ b^{(t+1)},{\mathbf{f}}^{(t+1)}, {\psi}^{(t+1)} \right\}.$
We show that we can train a deep neural network to achieves high accuracy in predicting the optimal communication link and serving beam.

\section{Proposed Deep Learning Based Solution}\label{sec:DL}
As the deep learning solution, we propose a recurrent neural network based on GRUs. Such networks have proven to be effective in learning sequence dependency, especially long sequences \cite{DLBook}. The proposed architecture is shown in Figure \ref{fig:deep_arch}. It has an input preparation stage that embeds the dual modality inputs, i.e., beam and location sequences, and then concatenates them to form a sequence of high dimensional vectors. This sequence is fed to a multi-layer GRU network. The network learns the sequential relation between the inputs and outputs a summary feature vector $\mathbf z$, which is passed to a classifier layer. This classifier is customized to predict either the communication-link or the best beam. In both cases, the classifier implements a fully-connected layer followed by a softmax layer.

\begin{figure*}[t]
	\centering
	\includegraphics[width=0.8\linewidth]{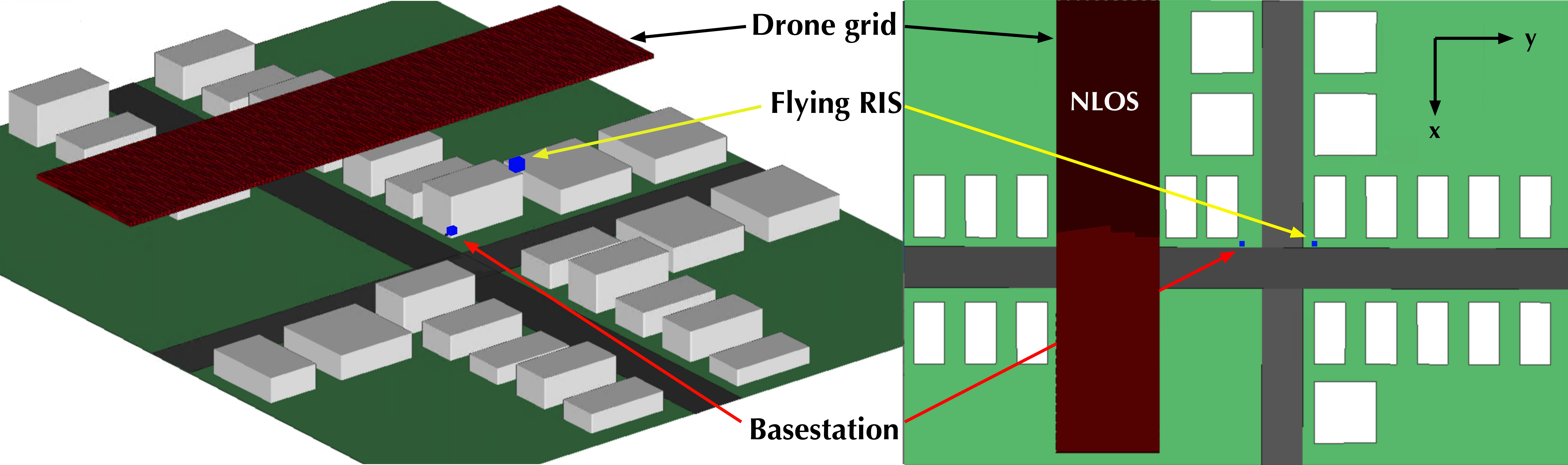}
	\caption{Outdoor drone-based scenario. \textbf{Left} is a perspective view, and \textbf{right} is a top-view.}
	\label{fig:scenario}
\end{figure*}

To ensure efficient training of the deep learning network, the inputs of the network go through a preparation stage, in which the inputs are standardized to have zero mean and unit variance \cite{EffBProp}. For the beam indices, an embedding layer is used to transform each beam value to an $a$-dimensional  Gaussian vector, drawn from a Gaussian distribution of zero mean and unit variance Note $a$ is the length of the embedding vector and it is a design parameter. In the simulation results, we use $a=50$. 
On the other hand, the locations are standardized by subtracting the mean (x,y,z) vector and dividing each dimension by its standard deviation. These statistics are computed from the training dataset, which will be described in Section \ref{sec:exp_setup}. Then, the standardized location and embedded vector of the beam index are concatenated forming a new $(3+a)$-dimensional vector.

\section{Dataset for Training and Testing of the Proposed Deep Learning  Solution}\label{sec:exp_setup}

To evaluate the proposed solution, a dataset representing a terahertz system, such as that described in Section~\ref{sec:sys_model}, needs to be developed. We use the DeepMIMO data-generation framework \cite{DeepMIMO} to build the scenario and develop that dataset. The scenario depicted in Figure \ref{fig:scenario} represents a downtown city intersection with a THz base station placed 6 meters high and a flying RIS hovering 80 meters above the ground. It also has a 3D drone grid with its base set 40 meters high. The 3D grid consists of 4 parallel 2D grids at 4 different heights. The spacing between the points of the grid is 0.81 meter along x- and y-axes and 0.8 meter along the z-axis. The intersection is surrounded by several buildings with same base area and different heights. In addition, each drone moves in these trajectories according to the random waypoint model. More details can be found in \cite{DeepMIMO_web}.

\begin{table}[h]
\centering
\caption{Parameters for data generation.}
\begin{tabular}{c | c | c} 
 \hline
 Parameter & BS & RIS \\ 
 \hline\hline
  Active BS & 1 & 2 \\ 
  Active user first & 1 & 1 \\
  Active user last & 496 & 496 \\
  Number of antennas (x,y,z) & (64,1,1) & (256,1,1) \\
  Antenna spacing & 0.5 & 0.5 \\
  Center Frequency &200 GHz &200GHz\\
  Bandwidth & 1 GHz & 1 GHz \\
  Number of OFDM subcarriers & 512 & 512 \\
  OFDM sampling factor & 1 & 1 \\
  OFDM limit & 1 & 1 \\
  Number of paths & 1 & 1 \\
  
 \hline
\end{tabular}
\label{table:hyper}
\end{table}

Using the DeepMIMO generation scripts, a seed dataset is generated and processed to construct the development dataset. The seed dataset has all the channels between every drone position in the 3D grid and both the RIS and the base station as well as the channel between the RIS and the base station. The DeepMIMO data generation parameters are listed in Table \ref{table:hyper}. With the channels in the seed dataset, a development dataset is built. It consists of 10-step sequences representing different drone trajectories. Each step in a sequence has a tuple of drone information, i.e., LOS status with the base station, LOS status with the RIS, best beamforming vector used by the base station, best beamforming vector used by the RIS, and the (x,y,z)-coordinates of the drone. A drone trajectory is formed by moving 10 steps in the 3D drone grid. Each step is taken randomly along one dimension and it is governed by the following probabilities: 0.2 along y-axis, 0.2 along z-axis, and finally 0.8 along +x-axis (a drone never moves backwards on x-axis). The development dataset has a little over 160 thousand sequences (data samples) of 10-step length. This data is shuffled and split $70\%-30\%$ to form the training and validation datasets.

\section{Numerical Results}\label{sec:exp_results}
The performance of the proposed model is evaluated with the development dataset described above. The next two subsections discuss the network architecture, its training procedure, and finally its evaluation results.

\subsection{Network Architecture and Training}
The proposed deep learning algorithm in Figure \ref{fig:deep_arch} is experimentally designed to achieve a good generalization performance. The hyperparameters of the deep learning network is summarized in Table~\ref{table:hyper1}. 
\begin{table}[h]
\centering
\caption{Deep learning network hyperparameters.}
\begin{tabular}{c | c } 
 \hline
 Hyperparameter & Value \\ 
 \hline\hline
  Number of GRU layers & 2  \\ 
  GRU unit dimension & 20 \\
  Percentage of dropout & 20\% \\
  Beam embedding space dimension & 50 \\
  Classifier layer dimension & 256\\
  Training algorithm & Adam  \\
  Number of training epochs & 100  \\
 \hline
\end{tabular}
\label{table:hyper1}
\end{table}

Through a sequence of experiments, the best performing architecture is found to have 2 layers of GRU units separated with a $20\%$ dropout layer. Each GRU unit has a hidden state with 20 dimensions and sees a sequence of length 7 (a drone trajectory). The beam embedding space has a dimensionality of $a = 50$. The output of the last GRU layer (at the 7-th step) is fed to classifier layer with 256 dimensions (number of classes). This is the result of $\max\{|\mathcal F|, |\mathcal P|\}$. The architecture is trained using Adam optimizer with a learning rate of $1\times 10^{-3}$ and for 100 epochs.

\begin{figure}[h]
	\centering
	\includegraphics[width=\linewidth]{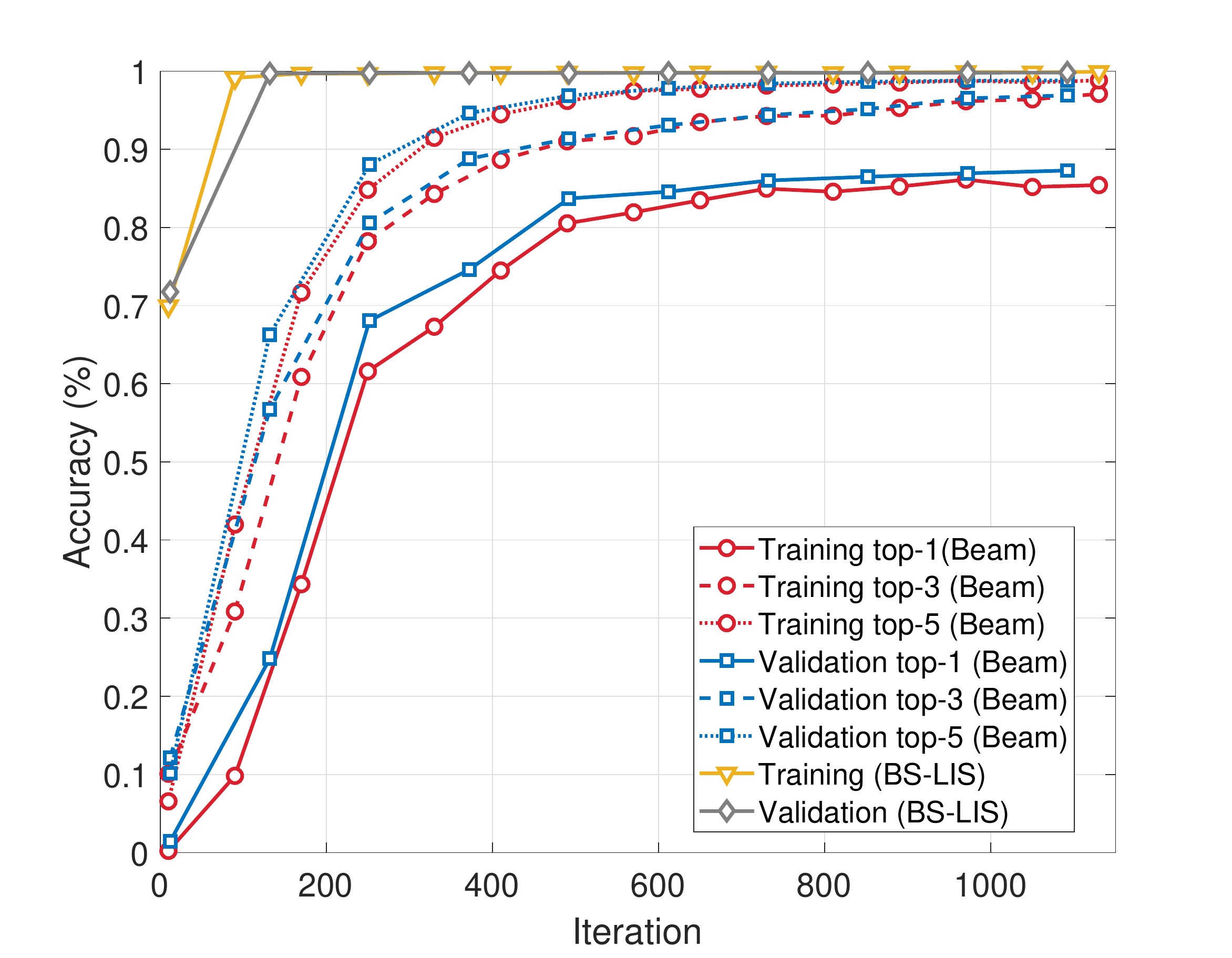}
	\caption{Prediction accuracy of the proposed network architecture versus training iteration. The figure depicts the performance on both tasks: beam and communication-link predictions.}
	\label{fig:accs}
\end{figure}

\begin{figure*}[t]
  \centering
  \includegraphics[width=0.95\linewidth]{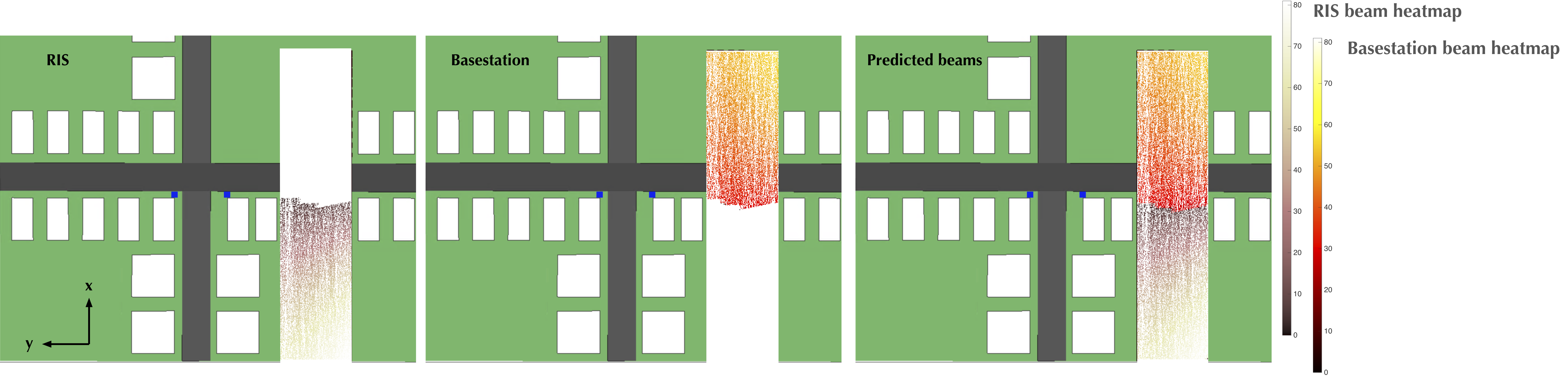}
  \caption{Beam distribution over the x-y plane of the drone grid. \textbf{Left image} shows groundtruth beams used by the RIS. \textbf{Center image} shows groundtruth beams used by the base station. \textbf{Right image} shows the predicted beams of both RIS and base station as given by the proposed network. All beams are obtained from the validation set.}
  \label{fig:heatmap}
\end{figure*}

\begin{figure}[ht]
	\centering
	\includegraphics[width=8.5 cm,height=7cm,angle=0]{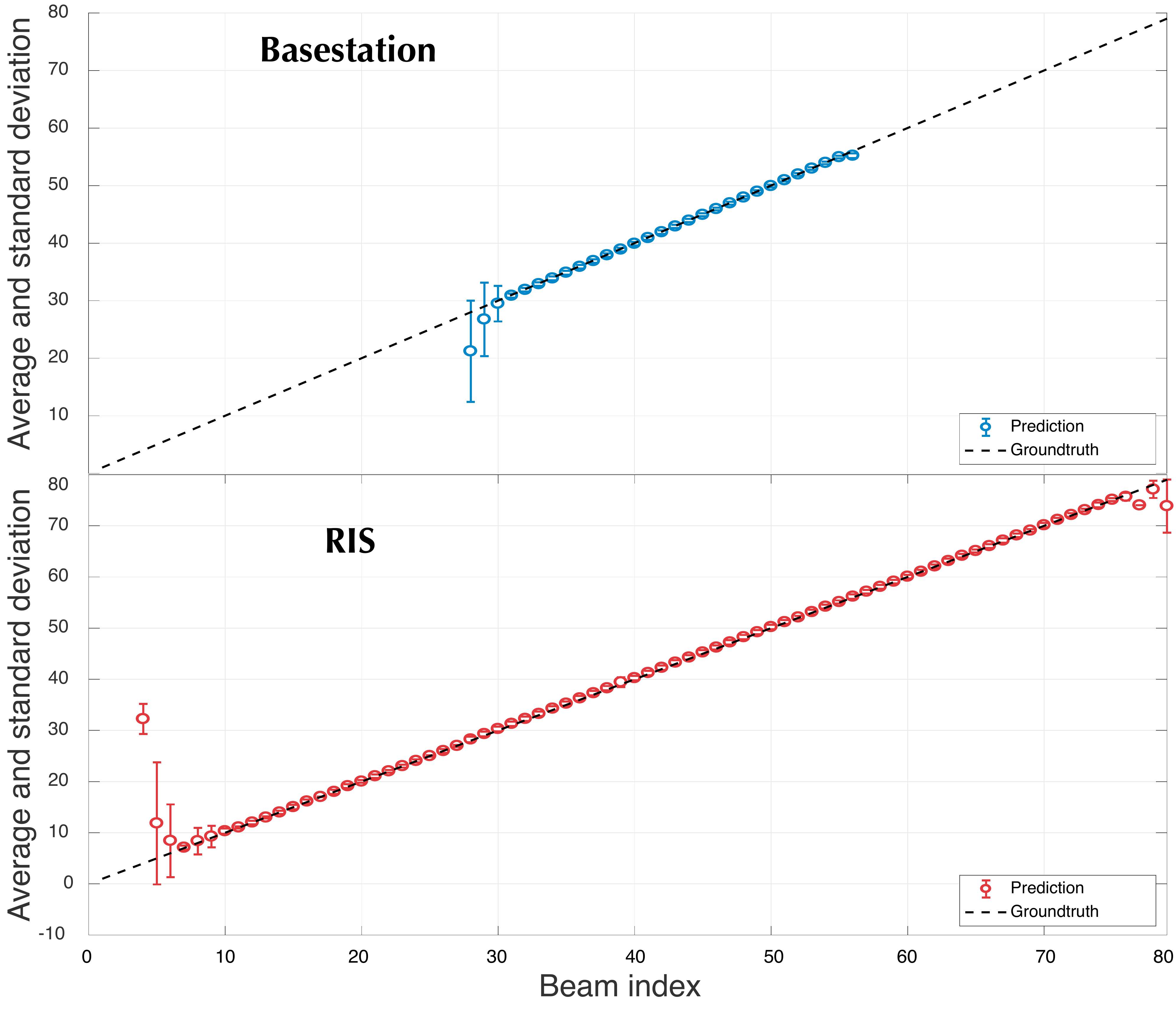}
	\caption{Per-beam prediction statistics (average and standard error) for both the base station and RIS codebooks.}
	\label{fig:stats}
\end{figure}
\subsection{Performance Evaluation}
The proposed architecture has two tasks: (i) predicting the future beam, i.e., beam to be used at the 8-th step, and (ii) predicting communication link, i.e., direct connection to the base station or RIS-assisted connection with the base station. Figure \ref{fig:accs} shows the training and validation performance of the architecture for both tasks. For beam prediction, this is quantified by plotting the top-1, top-3, and top-5 accuracies \cite{Sub6PredmmW} versus training iterations while for connection type, it is quantified using only top-1 since the prediction is binary. For beam prediction, top-1 accuracy shows that the architecture is likely to identify the correct beam with the accuracy of $\sim$$85\%$. This accuracy could be further improved with a little beam training by considering the top-3 and 5 beams predicted by the architecture. For instance, beam-training the top-3 predictions of the architecture provides a $11\%$ increase on top-1 prediction accuracy (from $\sim$$85\%$ to $\sim$$96\%$). This beam prediction performance is accompanied with a better prediction performance for the communication link. The same figure shows that the architecture achieves a near-perfect prediction for this task, hitting $\sim$$99.7\%$ accuracy. This performance is not surprising, though, due to the nature of the link blockage; the drones only experience link blockage with the base station when they are in the dark region illustrated in the top-view in Figure \ref{fig:scenario}. This region is a result of the corner building blocking the drone links to the base station. 

A deeper look at the performance of the architecture reveals two important conclusions: (i) the architecture has an almost even performance across all beams in $\mathcal F$ and $\mathcal P$, and (ii) it could achieve reliable ``drone handoff'' performance. The former is demonstrated in Figure \ref{fig:stats} where the average predicted value and its standard error are plotted against the groundtruth beam index for both the base station and RIS codebooks. The figure basically shows that for almost any choices of active beams (beams that are used in the scenario) at the base station or the RIS, the architecture on average predicts the right beam with almost zero standard error, e.g., whenever beam 50 is the correct choice for the 8-th step at both the base station and the RIS, the architecture predicts 50 with almost zero error. The figure also shows that the architecture only struggles when the groundtruth beam is close to the edge of the active beam set, i.e., beams 5-10 and beams 78-80 for the RIS and beams 28-31 for the base station. The reason for those anomalies (mis-predictions) could be attributed to where these beams are used; the left and middle images of Figure \ref{fig:heatmap} show the beam distribution over the drone x-y grid for both the base station and RIS. They clearly show that those anomalies occur close to the boundary between the LOS and NLOS regions with the base station (transition region). Finally, the right image in Figure \ref{fig:heatmap} demonstrates the second conclusion; not only the beam prediction is accurate, the communication-link prediction is also as good, which is a testament to how reliable drone hand-off is.

\section{Conclusion}\label{sec:conclusion}
In this paper, we considered the problem of proactive handoff and beam selection in a drone THz network employing RISs. To solve the problem, we proposed a novel deep learning based solution of GRU-based recurrent neural network that relies on the history of mobile user locations as well as the serving beams. Our results showed that the proposed deep learning solution yields high accuracy in proactive handoff and beam selection, and that with little beam training, the accuracy of the deep learning algorithm can be further improved. The increase in accuracy is up to $11\%$ when top-3 predictions is considered for beam prediction as compared to top-1 predictions. The improvement in the performance of beam prediction is also accompanied by an improvement of the communication-link prediction and consequently an improvement of the reliability in the drone hand-off. The results also show that the use of RISs is promising for proactive handoff and beam selection in a drone THz network.

\end{document}

%% file: input.tex
\usepackage{amsfonts}
\usepackage{times}
\usepackage{graphicx}
\usepackage{latexsym}
\usepackage{dsfont}
\usepackage{amssymb}
\usepackage{amsmath}
\usepackage{cite}
\usepackage{verbatim}

\def\bb0{{\mathbb{0}}}

\def\bb{{\mathbf{b}}}

\def\bff{{\mathbf{f}}}

\def\b0{{\mathbf{0}}}

\def\sf0{{\mathsf{0}}}

